\documentclass{kapproc} 

\usepackage{procps} 

\usepackage[dvips]{graphicx}

\upperandlowercase

 \let\footnote\savefootnote

\normallatexbib 

\begin{document}

\articletitle[The Phoenix Deep Survey]
{Extremely red galaxies in the Phoenix Deep Survey}

\author{A. M. Hopkins$^1$, J. Afonso, A. Georgakakis, M. Sullivan, B. Mobasher, L. E. Cram}

\affil{University of Pittsburgh, Department of Physics and Astronomy\\
3941 O'Hara St, Pittsburgh, PA 15206, USA\\
$^1$Hubble Fellow}


\begin{abstract}
The Phoenix Deep Survey (PDS) is a multiwavelength survey based on deep
1.4\,GHz radio observations used to identify a large sample of star forming
galaxies to $z=1$. Here we present an exploration of the evolutionary
constraints on the star-forming population imposed by the 1.4\,GHz source
counts, followed by an analysis of the average properties of extremely
red galaxies in the PDS, by using the ``stacking" technique.
\end{abstract}


\section{Introduction}
The study of galaxy evolution in recent years has included a strong
focus on the star formation properties of galaxies. Many of these studies are
based primarily on selection at ultraviolet (UV) and optical wavelengths,
and are known to be strongly affected by obscuration due to dust. It has
been shown that selection at these wavelengths results in samples of star
forming systems that omit a significant fraction of heavily obscured galaxies
\cite{2002ApJ...581..844S}. There have in addition been suggestions that the
most strongly star forming systems suffer the most obscuration
\cite{2003astro.ph..7175A, 2003astro.ph..6621H, 2002AXA...383..801B, 2001ApJ...558...72S, 2001AJ....122..288H}.
Using radio selection to construct a star forming galaxy sample
allows the detection of such systems, and the average obscuration in
this case is significantly higher than in optically
selected samples \cite{2003astro.ph..7175A,2003astro.ph..6621H}.

To identify a homogeneously selected catalogue of star forming
galaxies, unbiased by obscuration effects, and spanning
a broad redshift range ($0<z<1$), the Phoenix Deep Survey
(PDS) uses a deep 1.4\,GHz mosaic image, based on observations with the
Australia Telescope Compact Array. This has been used to construct one of the
largest existing deep 1.4\,GHz source catalogues \cite{2003AJ....125..465H}
(see http://www.atnf.csiro.au/people/ahopkins/phoenix/)
from which the star forming galaxy sample will be drawn. The PDS has already
been highly successful in providing a basis for several investigations
of the nature of star forming galaxies and their evolution (see references
in \cite{2003AJ....125..465H}). Throughout the present investigation
we assume a ($\Omega_M=0.3,\Omega_{\Lambda}=0.7,H_0=70$) cosmology.

\section{1.4\,GHz source counts and models}

\begin{figure}
\centerline{\rotatebox{-90}{\includegraphics[width=6.0cm]{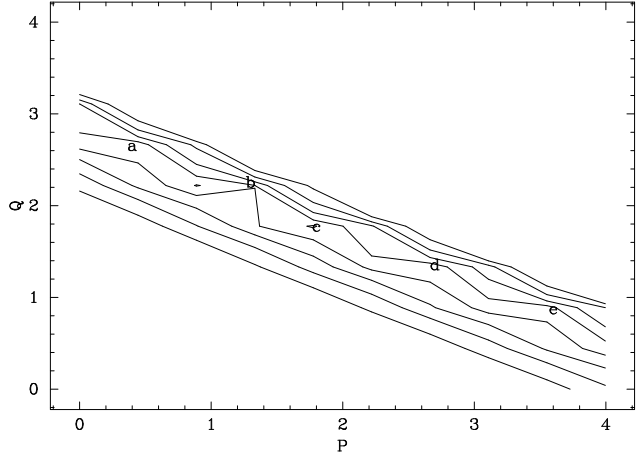}}}
\caption{Likelihood contours in the $(P,Q)$ plane, indicating the degeneracy
between rates of luminosity and density evolution for star-forming galaxies.
 \label{prob}}
\end{figure}

\begin{figure}
\centerline{\rotatebox{-90}{\includegraphics[width=6.0cm]{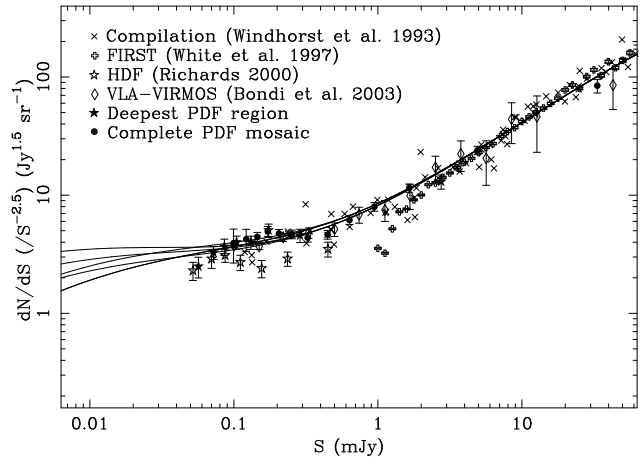}}}
\caption{Observed 1.4\,GHz source counts compared with model source count
predictions (solid lines). The models, while all predicting similar source
count distributions, span a broad range of combinations for the luminosity
and density evolution parameters.
 \label{srccnts}}
\end{figure}

The PDS 1.4\,GHz source counts have already been explored in detail
\cite{2003AJ....125..465H}. Here we explore source count models based on
local luminosity functions (LFs) and their assumed forms of evolution.
We assume an evolving LF for AGNs \cite{1990MNRAS.247...19D} (converted to
our assumed cosmology), since we are interested here in exploring the
properties of the star-forming (SF) galaxies. We then use a measured local
1.4\,GHz LF for SF galaxies \cite{2002MNRAS.329..227S}, and compare the
observed source counts with the model predictions subject to a range
of both luminosity evolution, $L(z)\propto (1+z)^Q$, and density evolution,
$\phi(z)\propto (1+z)^P$, for the SF population. The $\chi^2$ estimator for
each combination of $(P,Q)$, gives the statistical likelihood, and contours
showing the region of maximum likelihood in the $(P,Q)$ plane are shown
in Figure~\ref{prob}. There is a clear degeneracy, with maximum likelihood
occuring for any $(P,Q)$ satisfying $Q \approx 2.7 - 0.6\,P$ (note that
only positive $(P,Q)$ have been considered in this analysis). This sort of
degeneracy has also been shown in investigations of optically selected
galaxies \cite{1999ApJ...518..533L}.
To illustrate the effect of the degeneracy, Figure~\ref{srccnts}
shows the predicted source counts compared with the observations for five
$(P,Q)$ pairs, indicated by the positions a-e in Figure~\ref{prob}.
The redshift distributions predicted by these models, however, are
significantly different, and photometric redshifts are now being utilised
in order to differentiate between the different evolutionary models.

\section{Extremely red galaxies}

\begin{figure}
\centerline{\rotatebox{0}{\includegraphics[width=4.5cm]{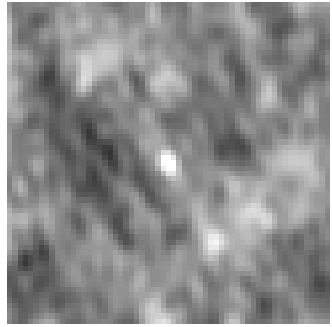}}}
\caption{Stacked image, incorporating 377 ERO candidates, from the PDS
radio mosaic. The image rms is about $1.1\,\mu$Jy, and the stacked
signal is detected at $\approx 6\,\sigma$ with a flux density of $6.5\,\mu$Jy.
 \label{ero_stack}}
\end{figure}

A sample of over 400 extremely red galaxies (ERGs) has been compiled
from PDS observations, using the colour criterion $R-K > 5$. Of these,
about 20 are detected at 1.4\,GHz, while the majority (over $90\%$)
remain undetected. In an effort to understand more about the properties
of these systems, the ``stacking analysis" technique often used at
X-ray wavelengths has been adopted. We extract sub-images from the
radio mosaic at the location of the non-detected ERGs, and construct
the weighted average of the sub-images (weighted by 1/rms$^2$, to maximise
the resulting signal-to-noise, since the radio mosaic has a varying noise
level over the image). Sub-images where low S/N emission ($>1.5\,\sigma$)
is present at the location of the non-detected source are excluded from
the stacking, in order to avoid biasing the stacking signal result by the
presence of a small number of low S/N sources. A further 22 (out of 399
candidate non-detected ERGs) were excluded in this manner, leaving 377
to contribute to the stacking signal. The stacked image, shown in
Figure~\ref{ero_stack}, has an rms noise of $1.1\,\mu$Jy, and a
$\approx6\,\sigma$ detection at $6.5\,\mu$Jy. While the actual redshift
distribution of these objects is unknown, a significant fraction of ERGs
appear to be dusty starbursts at $z \stackrel{>}{_{\sim}} 1$
\cite{2002ApJ...581..844S}. Assuming that the average redshift of these
sources is $z\approx 1$, the inferred average ERG 1.4\,GHz luminosity is
$3.5\times10^{22}\,$W\,Hz$^{-1}$. This corresponds to an average star
formation rate (assuming the ERGs are all star-forming systems) of 
$\approx 20\,M_{\odot}$yr$^{-1}$ (adopting the recent calibration of Bell 2003
\cite{2003ApJ...586..794B}; this value doubles if the calibration
of Condon 1992 \cite{1992ARA&A..30..575C} is used). Clearly these results
rely on a large number of assumptions, but they serve the useful purpose
of providing a preliminary estimate for the average properties of these
systems in order to support proposals for further, more detailed
investigation.

AMH acknowledges support provided by the National Aeronautics and Space
Administration (NASA) through Hubble Fellowship grant 
HST-HF-01140.01-A awarded by the Space Telescope Science Institute (STScI).
JA gratefully acknowledges the support from the Science and
Technology Foundation (FCT, Portugal) through the fellowship
BPD-5535-2001 and the research grant POCTI-FNU-43805-2001.


\begin{chapthebibliography}{}

\bibitem[1]{2003astro.ph..7175A} Afonso, J., Hopkins, A., Mobasher, B.,
\& Almeida, C.\ 2003, ApJ, (in press; astro-ph/0307175)
\bibitem[2]{2003ApJ...586..794B} Bell, E.~F.\ 2003, ApJ, 586, 794
\bibitem[3]{2002AXA...383..801B} Buat, V., Boselli, A.,
Gavazzi, G., \& Bonfanti, C.\ 2002, A\&A, 383, 801
\bibitem[4]{1992ARA&A..30..575C} Condon, J.~J.\ 1992, ARA\&A, 30, 575
\bibitem[5]{1990MNRAS.247...19D} Dunlop, J.~S.~\&
Peacock, J.~A.\ 1990, MNRAS, 247, 19
\bibitem[6]{2003AJ....125..465H} Hopkins, A.~M., Afonso, J., Chan, B.,
Cram, L.~E., Georgakakis, A., \& Mobasher, B.\ 2003, AJ, 125, 465
\bibitem[7]{2001AJ....122..288H} Hopkins, A.~M., Connolly, A.~J.,
Haarsma, D.~B., \& Cram, L.~E.\ 2001, AJ, 122, 288
\bibitem[8]{2003astro.ph..6621H} Hopkins, A.~M.~et al.\ 2003, ApJ,
(submitted; astro-ph/0306621)
\bibitem[9]{1999ApJ...518..533L} Lin, H., Yee, H.~K.~C.,
Carlberg, R.~G., Morris, S.~L., Sawicki, M., Patton, D.~R., Wirth, G., \&
Shepherd, C.~W.\ 1999, ApJ, 518, 533
\bibitem[10]{2002MNRAS.329..227S} Sadler, E.~M.~et al.\ 2002, MNRAS, 329, 227
\bibitem[11]{2002ApJ...581..844S} Smail, I., Owen, F.~N., Morrison, G.~E.,
Keel, W.~C., Ivison, R.~J., \& Ledlow, M.~J.\ 2002, ApJ, 581, 844
\bibitem[12]{2001ApJ...558...72S} Sullivan, M., Mobasher, B., Chan, B.,
Cram, L., Ellis, R., Treyer, M., \& Hopkins, A.\ 2001, ApJ, 558, 72

\end{chapthebibliography}
\end{document}